\documentclass[12pt]{iopart}
\usepackage{graphicx}

\begin{document}
\def \ee {\varepsilon}
\thispagestyle{empty}
\title[]{
Stronger constraints on non-Newtonian gravity
from the Casimir effect
}

\author{
V~M~Mostepanenko,${}^{1,2}$ R~S~Decca,${}^3$ E~Fischbach,${}^4$
G~L~Klimchitskaya,${}^{1,5}$ D~E~Krause${}^{6,4}$
and
D~L\'{o}pez${}^{7}$
}

\address{${}^1$Center of Theoretical Studies and Institute for Theoretical 
Physics, Leipzig University,
D-04009, Leipzig, Germany}

\address{${}^2$
Noncommercial Partnership  ``Scientific Instruments'', 
Tverskaya St. 11, Moscow, 103905, Russia}

\address{${}^3$
Department of Physics, Indiana University-Purdue
University Indianapolis, Indianapolis, Indiana 46202, USA}

\address{${}^4$
Department of Physics, Purdue University, West Lafayette, Indiana
47907, USA}
 
\address{$^5$North-West Technical University, Millionnaya St. 5,
St.Petersburg, 191065, Russia}

\address{${}^6$
Physics Department, Wabash College, Crawfordsville, Indiana 47933,
USA}

\address{${}^7$
Bell Laboratories, Lucent Technologies, Murray Hill,
New Jersey 07974, USA}

\begin{abstract}
We review new constraints on the Yukawa-type corrections to
Newtonian gravity obtained recently from gravitational experiments
and from the measurements of the Casimir force. Special attention is
paid to the constraints following from the most precise dynamic
determination of the Casimir pressure between the two parallel plates
by means of a micromechanical torsional oscillator. The possibility
of setting limits on the predictions of chameleon 
field theories using the results
of gravitational experiments and Casimir force measurements is discussed.
\end{abstract}

\section{Introduction}

During the last ten years hypothetical long-range interactions coexisting
with Newtonian gravity have received much attention.
There are serious reasons why the existence of such interactions is very
probable. Many extensions of the standard model predict light elementary
particles, such as axions, scalar axions, dilatons, graviphotons, etc. 
The exchange of such particles between two atoms with masses $M_1$ and
$M_2$ at a separation $r$ results in an attractive or repulsive force
described by the effective Yukawa-type potential which is added to the
usual gravitational potential \cite{1}
\begin{equation}
V_{\rm Yu}(r)=-\frac{GM_1M_2}{r}\left(1+\alpha{\rm e}^{-r/\lambda}\right).
\label{eq1}
\end{equation}
\noindent
Here, $G$ is the gravitational constant, $\alpha$ is the interaction
constant of a hypothetical interaction relative to gravity,
and $\lambda$ is the interaction range ($\lambda=m^{-1}$ where $m$
is the mass of a hypothetical particle). Exchange of massless particles
(neutrinos or arions, for instance) leads to the power-type corrections
to Newtonian gravity with different powers \cite{2,3}
\begin{equation}
V_{l}(r)=-\frac{GM_1M_2}{r}\left[1+\Lambda_l
\left(\frac{r_0}{r}\right)^{l-1}\right],
\label{eq2}
\end{equation}
\noindent
where $l=1,\,2,\,3,\ldots\,$,
$\Lambda_l$ is the interaction constant and the arbitrary parameter
$r_0=10^{-15}\,$m is introduced for preserving the proper dimensionality
of the potential.

Another theoretical scheme that predicts corrections to Newton's 
gravitational law is extra-dimensional physics with low compactification energy
$M_{\rm Pl}^{(N)}=1/G_{4+n}^{1/(2+n)}\sim 1\,$TeV, where $G_{4+n}$ is
the gravitational constant in $N=(4+n)$-dimensional space-time and $n$
is the number of extra spatial dimensions. This energy should be compared
with the usual Planck energy $M_{\rm Pl}=1/\sqrt{G}\sim 10^{19}\,$GeV.
The size of the compact extra dimensions is given by \cite{4,5}
\begin{equation}
 R_n\sim\frac{1}{M_{\rm Pl}^{(N)}}\,\left(
\frac{M_{\rm Pl}}{M_{\rm Pl}^{(N)}}\right)^{2/n}\sim
10^{\frac{32}{n}-17}\,\mbox{cm}.
\label{eq3}
\end{equation}
\noindent
Under the condition that $r\gg R_n$, low energy compactification schemes 
predict  Yukawa-type corrections to Newtonian  gravity, as in
(\ref{eq1}), with $\lambda\sim R_n$ \cite{6,7}. For $n=1$ it follows
that $R_1\sim 10^{15}\,$cm which is excluded by gravity tests in the
solar system \cite{1}. However, for $n=2$ and 3 the sizes of predicted
extra dimensions are $R_2\sim 1\,$mm and $R_3\sim 5\,$nm, i.e., 
the very ones presently tested in the laboratory experiments of Cavendish-
and E\"{o}tvos-type and in the measurements of the Casimir force.

Another proposed scheme deals with noncompact but warped extra dimensions
\cite{8} and this leads to power-type correction to Newtonian
gravity, as in (\ref{eq2}), with $l=3$.

Recently one more extension of the standard model, the so-called
{\it chameleon field theory}, became very popular. As with many other
extensions of the standard model, this theory introduces one or more
scalar fields. A specific feature of these fields is that their masses depend 
on the local background matter density and they can couple directly to
matter with gravitational strenth \cite{9,10}. The chameleon scalar field,
if it really exists in nature, leads to an additional chameleon force
acting  between two nearby macrobodies. The functional dependence of
this force on the separation distance is rather complicated and it depends
on the specific form of the potential of the chameleon field. Typically
the chameleon force behaves as an inverse power of distance between the
two macrobodies but other asymptotic regimes are also possible \cite{11,11a}.

All of the above predictions made in physics beyond the standard model can be
tested using gravitational experiments and measurements of the
Casimir force. In this paper we briefly review the progress achieved
in the strengthening of constraints on non-Newtonian gravity during the
two years passed after the QFEXT05 
conference in Barcelona. In Section 2 new
constraints obtained from precise gravitational measurements are presented.
Section 3 is devoted to the constraints following from the most precise
determination of the Casimir pressure between the two parallel plates using
a micromechanical torsional oscillator. Section 4 contains our conclusions
and prospects. We use units with $c=\hbar=1$.

\section{Constraints following from gravitational experiments}

Gravitational experiments of the E\"{o}tvos- and Cavendish-type have a long
history. They have been considered as the most precise physical
experiments over many years. E\"{o}tvos-type experiments measure limits
on the relative difference in accelerations imparted by  the Earth,
Sun or some laboratory attractor to various substances of the same mass. 
In Cavendish-type experiments, limits on the deviations from the
force-distance dependence of $1/r^2$ in the Newton gravitational law are
measured. The results of both types of experiments can be used to constrain
the interaction constants ($\lambda,\alpha$) and $\Lambda_l$ in the
interaction potentials (\ref{eq1}), (\ref{eq2}) \cite{12}. In figure 1 we 
present the strongest constraints obtained from gravitational experiments
on the parameters of a Yukawa-type hypothetical interaction ($\lambda,\alpha$).
Lines 1,\,2 and 3 are obtained from the experiments of papers
\cite{13,14,15}, respectively. Permitted regions on ($\lambda,\alpha$)-plane
lie beneath the lines.

\begin{figure*}[b]
\vspace*{-14.cm}
\hspace*{2cm}\includegraphics{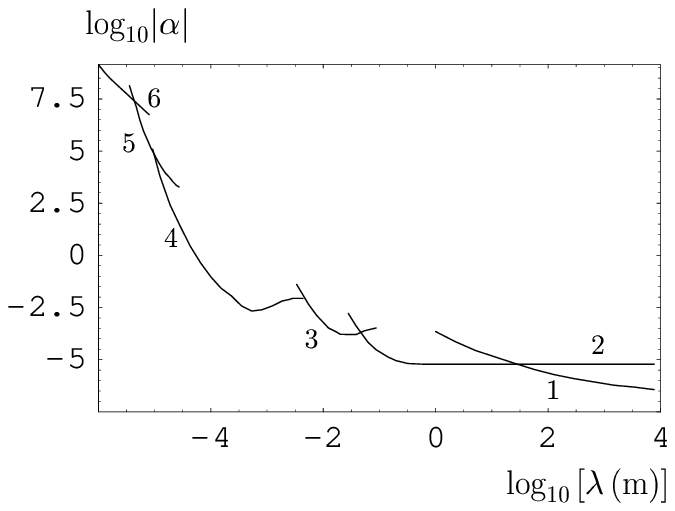}
\vspace*{-11.8cm}
\caption{
Strongest constraints on Yukawa-type corrections to Newton's
gravitational law following from different gravitational experiments
(lines 1--5) and the measurement of the Casimir force (line 6).
Permitted regions in the ($\lambda,\alpha$)-plane
lie beneath the lines (see text for further discussion).
}
\end{figure*}
During the last two years, constraints on the parameters of
Yukawa-type interactions were strengthened in two new important
gravitational experiments \cite{16,17}. In \cite{16}
a micromechanical cantilever was used as the force sensor, and its 
displacement was measured interferometrically to find constraints
on the Yukawa-type deviations from Newtonian gravity. The results
of this experiment are shown as line 5 in figure 1. The largest
improvement over previous results obtained in \cite{18,19} is by
almost a factor of 10 at $\lambda\approx 20\,\mu$m. The experiment \cite{17}
sets stronger constraints on deviations from Newton's inverse-square
law using a torsion-pendulum detector suspended above an attractor that
was rotated with an uniform angular velocity. The resulting constraints
are shown as line 4 in figure 1. The previously known constraints in
this region found in \cite{16,18,20} are improved by a factor of up to
100 by the results of this experiment (line 6 shows constraints \cite{21}
following from the measurement of the Casimir force using a torsion
pendulum \cite{22}).
This shows that gravitational experiments have considerable potential in
further strengthening the constraints on Yukawa-type corrections to
Newtonian gravity for $\lambda$ larger than a few micrometers. 
At the same time, constraints on the parameters $\Lambda_l$ of power-type
interactions have not been strengthened during the last time
(see \cite{23} for the list of the strongest constraints).

The gravitational experiments have  the potential to constrain some
predictions of chameleon theories. The predictions of these theories
 depend,
however, on where the gravitational experiment is performed. If it is
performed in the low-density vacuum of space, the magnitude of the
chameleon force might be larger than if it is performed in the relatively
high-density environment of a laboratory.
According to \cite{11a}, the experiment \cite{17} could detect or rule out 
the existence of chameleon fields with some natural values of parameters,
provided it is designed to do so. In particular, the role of electrostatic
forces should be eliminated without using a metallic sheet between the
attractor and pendulum. Such a sheet is used presently, but 
it plays the crucial
role when testing for chameleon fields.

\section{Constraints on the Yukawa interaction from Casimir
force measurements}

Measurements of the Casimir force are now generally recognized as the
source of constraints on Yukawa-type corrections to Newtonian gravity.
During the last few years significant progress has been made in  
increasing the experimental precision and in the 
comparison of the measurement
data with theory at a given confidence level \cite{24,25}. This 
has permitted us to
obtain constraints of the same reliability as those following from the
gravitational experiments. Typically measurements of the Casimir force 
allow one to obtain constraints on hypothetical interactions with a shorter
interaction range than gravitational experiments. Thus, both types
of experiments play a supplementary role in constraining the hypothetical 
interactions of Yukawa-type.

The basic idea on how the Casimir force measurements can be used for
constraining hypothetical long-range interactions is the following.
The hypothetical interaction of Yukawa-type (\ref{eq1}) leads to some
additional force in the experimental configuration where the Casimir
force is measured. This additional force depends on unknown parameters
$\alpha$ and $\lambda$. If the measurement data for the Casimir force
are consistent with respective theory within some confidence interval,
the hypothetical force must be sufficiently small. This imposes
constraints on $\alpha$ and $\lambda$.

Here we present constraints on Yukawa-type corrections to Newton's
gravitational law following from recent dynamic determinations of the
Casimir pressure between the two gold coated parallel plates by means
of a micromechanical torsional oscillator \cite{26,27}. In this
experiment a large sphere is oscillating above a plate with
the natural frequency of the oscillator and the frequency shift due to the 
Casimir force is measured. By means of the proximity force approximation, 
the frequency shift is recalculated into the equivalent Casimir
pressure between two plates. The experiment under discussion
is the first measurement of the Casimir force of metrological quality
in the sense that the stochastic experimental error is much smaller
than  the systematic error. As a result, it is the systematic error alone that
determines the total experimental error over the entire measurement range. 
The total experimental error of the Casimir pressure measurements
determined at a 95\% confidence level varies from 0.19\% of the measured
pressure at a separation $a=162\,$nm, to 0.9\% at $a=400\,$nm,
and to 9.0\% at $a=746\,$nm. The description of the experimental setup,
the measurement procedure, and of the comparison of data with theory
can be found in \cite{26,27}.

Constraints on the Yukawa-type hypothetical interaction are obtained from 
the measure of agreement between the experimental data and theory. This 
can be quantified as a 95\% confidence band 
$[-\tilde\Xi(a),\tilde\Xi(a)]$ containing no less than 95\% of all
differences $P^{\rm th}(a)-\bar{P}^{\rm exp}(a)$ in the measurement range
from 180 to 746\,nm, where $P^{\rm th}(a)$ is the calculated value of the
Casimir pressure at a separation $a$ and $\bar{P}^{\rm exp}(a)$ is the
mean measured value at the same separation.
The function $\tilde\Xi(a)$ is determined by both the experimental errors
discussed above and the theoretical errors in the calculation of the
Casimir pressure. In \cite{27} $\tilde\Xi(a)$ was determined in a conservative 
way, such that the confidence band $[-\tilde\Xi(a),\tilde\Xi(a)]$ includes
not only 95\%, but 100\% of differences between the theoretical and
mean experimental Casimir pressures (note that data from
the shortest separations
between 162 and 180\,nm were not used for obtaining constraints). For
example, at typical separations $a=180$, 200, 250, 300, 350, 400 and
450\,nm, the half-widths of the confidence band are equal to
$\tilde\Xi(a)=4.80$, 3.30, 1.52, 0.84, 0.57, 0.45, and 0.40\,mPa,
respectively. From this, the magnitude of the hypothetical pressure
can be constrained from the inequality
\begin{equation}
|P^{\rm hyp}(a)|\leq\tilde\Xi(a).
\label{eq4}
\end{equation}
\noindent
The constraints obtained from (\ref{eq4}) are characterized by the same
confidence as $\tilde\Xi(a)$, i.e., by the 95\% confidence level.

The hypothetical pressure resulting from the potential (\ref{eq1})
can be obtained by the integration of (\ref{eq1}) over the volumes of
the plates, and subsequent negative differentiation with respect to $a$.
In so doing the contribution from the gravitational interaction [the first
term in (\ref{eq1})] can be neglected \cite{29,30}. In this dynamic
experiment one plate is effective and has the same layer structure as a large
oscillating sphere of radius $R$. Thus, it is made of sapphire of
density $\rho_s=4.1\,\mbox{g/cm}^3$ coated with a layer of Cr of
density $\rho_c=7.14\,\mbox{g/cm}^3$ and thickness $\Delta_c=10\,$nm, and
then with an external layer of gold of thickness 
$\Delta_g^{\!(s)}=180\,$nm and density $\rho_g=19.28\,\mbox{g/cm}^3$.
The other (real) plate is made of Si of thickness $L=3.5\,\mu$m and
density $\rho_{Si}=2.33\,\mbox{g/cm}^3$. It was first coated with a layer
of Cr of $\Delta_c=10\,$nm thickness and then with a layer of gold of
$\Delta_g^{\!(p)}=210\,$nm thickness. Note that both sapphire and Si
can be considered as infinitely thick.
Under the conditions $a,\,\lambda\ll R$, the equivalent Yukawa pressure
between the two parallel plates with the above layer structure is
given by \cite{21,31}
\begin{eqnarray}
&&
P^{\rm hyp}(a)=-2\pi G\alpha\lambda^2{\rm e}^{-a/\lambda}
\label{eq5} \\
&&\phantom{aaaa}
\times\left[\rho_g-(\rho_g-\rho_c){\rm e}^{-\Delta_g^{\!(s)}/\lambda}-
(\rho_c-\rho_s){\rm e}^{-(\Delta_g^{\!(s)}+\Delta_c)/\lambda}\right]
\nonumber \\
&&\phantom{aaaa}
\times\left[\rho_g-(\rho_g-\rho_c){\rm e}^{-\Delta_g^{\!(p)}/\lambda}-
(\rho_c-\rho_{Si}){\rm e}^{-(\Delta_g^{\!(p)}+\Delta_c)/\lambda}\right].
\nonumber
\end{eqnarray}

We have substituted (\ref{eq5}) in (\ref{eq4}) and found constraints on
the parameters of Yukawa interaction $\lambda,\,\alpha$ at different
separations $a$. The strongest constraints are shown in figure 2 by line 1.
For different $\lambda$, the strongest constraints are obtained at
different separations $a$. As an example, for
$10\,\mbox{nm}<\lambda<56\,$nm, the comparison of experiment with theory
at a separation of $a=180\,$nm leads to the strongest constraints.
\begin{figure*}[t]
\vspace*{-12.5cm}
\hspace*{2cm}\includegraphics{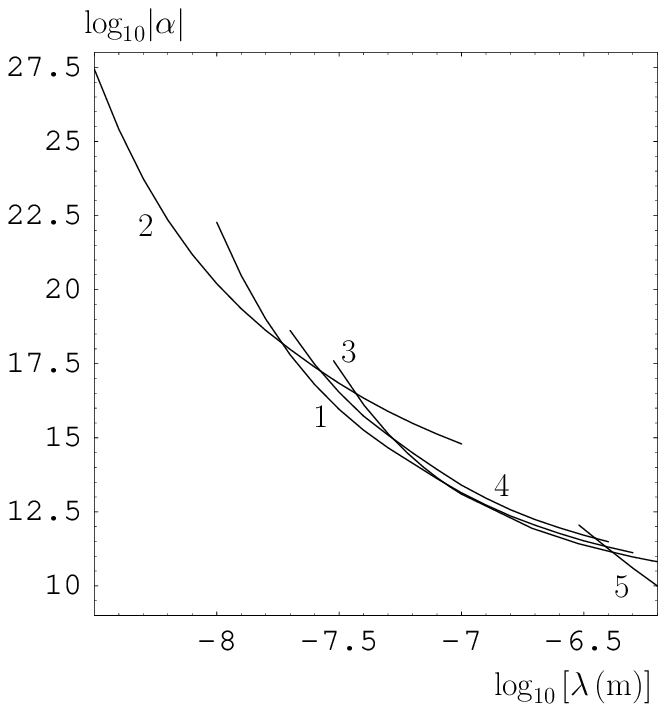}
\vspace*{-10.2cm}
\caption{
Constraints on Yukawa-type corrections to Newton's
gravitational law following from different  
measurements of the Casimir force (see text for further discussion).
Permitted regions on ($\lambda,\alpha$)-plane
lie beneath lines 1--5.
}
\end{figure*}
For illustration, constraints from earlier experiments are also shown in
figure 2. Line 2 follows from the short-separation measurement of the
Casimir force between a sphere and a plate using an atomic force
microscope \cite{32,33}. Note that for the first time the constraints
following from this experiment were obtained in \cite{29} at an
undetermined confidence level. Here, line 2 is recalculated at a 95\%
confidence level using the improved procedure for the comparison
of the Casimir force measurements with theory, as described in
\cite{24,25}. Line 3 was obtained in \cite{34} using the isoelectronic
technique. Line 4 follows from the previous experiment of the dynamic
determination of the Casimir pressure by means of a micromechanical
torsional oscillator \cite{24}. Line 5 was obtained \cite{21} from
the measurement of the Casimir force using a torsion pendulum
\cite{22}. It is, in fact, a continuation of the line labeled 6 in figure 1.

As is seen in figure 2, the constraints represented by line 1 are the
strongest ones within the interaction range from 20 to 86\,nm.
The largest improvement of previously obtained results is by a
factor of 4.4 at $a=26\,$nm.

It is of interest to consider constraints on the predictions of
chameleon field theories which follow from the measurements of the Casimir
force. The typical potential of the chameleon field $\phi$ can be chosen in
the form \cite{11}
\begin{equation}
V(\phi)=\Gamma_0^4\left(1+\frac{\Gamma^n}{\phi^n}\right),
\label{eq6}
\end{equation}
\noindent
where $n$ can be both positive and negative, and $\Gamma_0, \Gamma$ are
some constants. To fit data for the acceleration of the Universe, one 
requires $\Gamma_0\approx 2.4\times 10^{-3}\,$eV.
The hypothetical pressure $P_{\phi}$ arising between the two parallel
plates in chameleon theories with potential (\ref{eq6}) and
$\Gamma=\Gamma_0$ was calculated in \cite{11}. It was shown that for
$n>0$ the most precise experimental results of \cite{26,27} do not
impose any constraints on predictions of chameleon theories. The current
limits in figure 2 should be strengthened by at least two orders
of magnitude in order for constraints on chameleon theories with
$n>0$ to be obtained. At the same time 
the experimental data of \cite{26,27}
rule out the chameleon theories with $n=-4$ and $n=-6$ \cite{11}. Future
Casimir force measurements at large separations can be used to obtain
more stringent constraints on the predictions of chameleon 
field theories.

\section{Conclusions and discussion}

As was discussed above, during the last two years new important gravitational
experiments and Casimir force measurements have been performed which lead to 
stronger constraints on Yukawa-type corrections to Newtonian gravity.
The stronger constraints obtained from the gravitational measurements
are related to the interaction range from about $4\,\mu$m to
$4000\,\mu$m. Constraints strengthened from the measurement of the Casimir
pressure between two parallel plates are related to shorter interaction
scales from 20 to 86\,nm.
Thus, both experimental approaches used to strengthen constraints on
non-Newtonian gravity are complementary.

One important innovation introduced in the measurements of the Casimir force
during the last years is the increased experimental precision that
permitted us
to obtain data of metrological quality, where stochastic errors are
much below the systematic errors. Another innovation is the use of
rigorous statistical procedures for data processing and for the
comparison of experiment with theory, which allowed 
us to obtain constraints
at a fixed high confidence level. Taken together, these innovations
significantly increased the reliability of the resulting constraints on
non-Newtonian gravity, bringing them closer to the previously achieved high
reliability constraints following from the gravitational experiments.

An interesting new direction, which came into being recently, is the
application of Casimir force measurements to obtain constraints on
the predictions of chameleon field theories. First results in this
direction have been already obtained (see above).
New experiments planned for the near future promise to provide much more
information on this subject, especially if the chameleon theories
become more certain than they presently are. In this respect a 
more precise laboratory technique for probing small forces
in submicrometer range (see, e.g., \cite{35})
is of high promise.

All this permits us
to conclude that relatively inexpensive laboratory measurements
of the Casimir force continue to have great potential to obtain new
information on elementary particles and fundamental interactions.

\section*{Acknowledgments}
VMM and GLK acknowledge support from DFG 
Grant No.~436\,RUS\,113/789/0--3.
They are
grateful to the Center of Theoretical Studies and Institute
for Theoretical Physics, Leipzig University for kind
hospitality.
RSD acknowledges NSF support through Grant No. CCF--0508239. 
EF was supported
in part by DOE under Grant No. DE--AC02--76ER071428.

\end{document}